\documentclass[prl,twocolumn,nofootinbib]{revtex4}

\usepackage{graphicx}
\usepackage{amsmath}
\usepackage{color}
\usepackage{ulem}

\newcommand{\ket}[1]{|{#1}\rangle}
\newcommand{\bra}[1]{\langle {#1}|}

\newcommand{\be}{\begin{equation}}
\newcommand{\ee}{\end{equation}}

\begin{document}
\title{Nonadiabatic Diffraction of Matter Waves}

\author{J. Reeves, L. Krinner, M. Stewart, A. Pazmi\~{n}o, D. Schneble}

\affiliation{Department of Physics and Astronomy, Stony Brook University, Stony Brook, New York 11794-3800, USA}

\date{\today}

\begin{abstract}
Diffraction phenomena usually can be formulated in terms of a potential that induces the redistribution of a wave's momentum. Using an atomic Bose-Einstein condensate coupled to the orbitals of a state-selective optical lattice, we investigate a hitherto unexplored nonadiabatic regime of diffraction in which no diffracting potential can be defined, and in which the adiabatic dressed states are strongly mixed. We show how, in the adiabatic limit, the observed coupling between internal and external dynamics gives way to standard Kapitza-Dirac diffraction of atomic matter waves. We demonstrate the utility of our scheme for atom interferometry and discuss prospects for studies of dissipative superfluid phenomena.
\end{abstract}


\maketitle

Diffraction, the bending of waves around obstacles, is one of the most fundamental and ubiquitous phenomena in optics, with a centuries-old history going back to the works of Grimaldi, Huygens, and Young on the wave nature of light. In the modern era it has led to the understanding of x-rays
\cite{Bragg1913,Laue1913}, has provided direct proof for the wave nature of particles
\cite{Davisson1927}, and today finds many applications in physics and materials science, ranging from electron, x-ray and neutron diffraction, to applications in atom optics \cite{Meystre2001,Cronin2009}.


Quite generally, diffraction is caused by a position-dependent potential with absorptive (imaginary) and/or dispersive (real) character. While the former includes material structures and diffraction gratings, an example for the latter is the ponderomotive potential exerted on electrons by an optical standing wave, as
originally suggested by Kapitza and Dirac \cite{Kapitza1933}. Kapitza-Dirac diffraction of electrons was first
demonstrated only fairly recently \cite{Freimund2001}, well after the first observation of an analogous effect on neutral atoms based on the ac Stark shift near an atomic resonance \cite{Gould1986}. It has also been applied to Bose-Einstein condensates \cite{Ovchinnikov1999,Stenger1999} and has become an important and often-used tool in atom interferometry and metrology \cite{Cronin2009}.

In this paper, we investigate a hitherto unexplored regime of diffraction, in which a diffracting potential cannot be defined.  In our experiment we diffract an atomic matter wave from a microwave-dressed optical lattice, with a diffractive dynamics that is qualitatively different from Kapitza-Dirac diffraction of dressed matter waves from a periodic optical potential. We note that the coherent mixing of states interacting with an external field is often used for the engineering of dressed potentials \cite{Ketterle1996,Zobay2001,Colombe2004,Cohen-Tannoudji2004,Lesanovsky2006,White2006,Hofferberth2006,Lundblad2008,Lin2009}.
Deviations from adiabaticity have previously been found to have deleterious effects on dressed-state lifetimes \cite{Yi2008,Lundblad2014}. In our experiment, we enter the strongly nonadiabatic regime, in which coherent Landau-Zener transitions of the atomic wavefunction between the adiabatic dressed potentials (driven by zero-point motion) induce a strong coupling between internal and external degrees of freedom, leading to a breakdown of the usual Born-Oppenheimer (adiabatic) approximation \cite{Dalibard2011}.

\begin{figure}[t]
\centering
\includegraphics[width=\columnwidth]{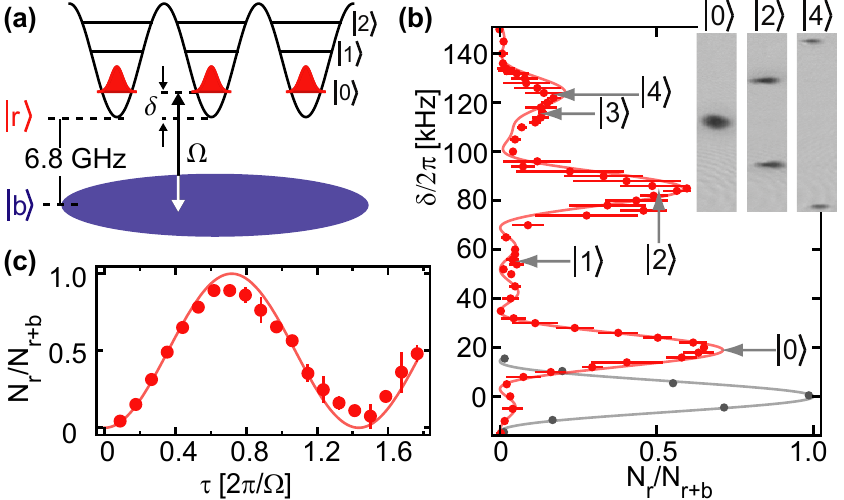}
\caption{(a)  Experimental scheme, shown for resonant coupling to the lowest orbital $\ket{n}$ ($n=0$) (see text).  (b) Fraction of atoms in the lattice after applying a pulse (Rabi frequency $\Omega/2\pi=8.1$kHz, duration $\tau=60\mu$s) with variable detuning $\delta$. The red solid line is the incoherent sum of predicted line shapes for resonances centered at detunings given by the $q=0$ level structure of the lattice; the pulse areas are $\pi\gamma_n$, where $\gamma_n=0.72, 0, 0.61, 0, 0.32$  for $n=0,1,2...$.  The light gray points and line refer to a reference measurement without lattice (pulse area $\pi$). The inset shows TOF absorption images after band-mapping. (c) Population dynamics in orbital $\ket{0}$ for resonant driving with $\Omega/2\pi=8.8$kHz and $\delta/2\pi=19.2$kHz. The curve is a resonant Rabi oscillation with frequency $\gamma_0 \Omega$. All error bars represent statistical standard deviations of at least three experimental iterations.} \label{fig_1}
\end{figure}

\begin{figure}
\centering
\includegraphics[width=\columnwidth]{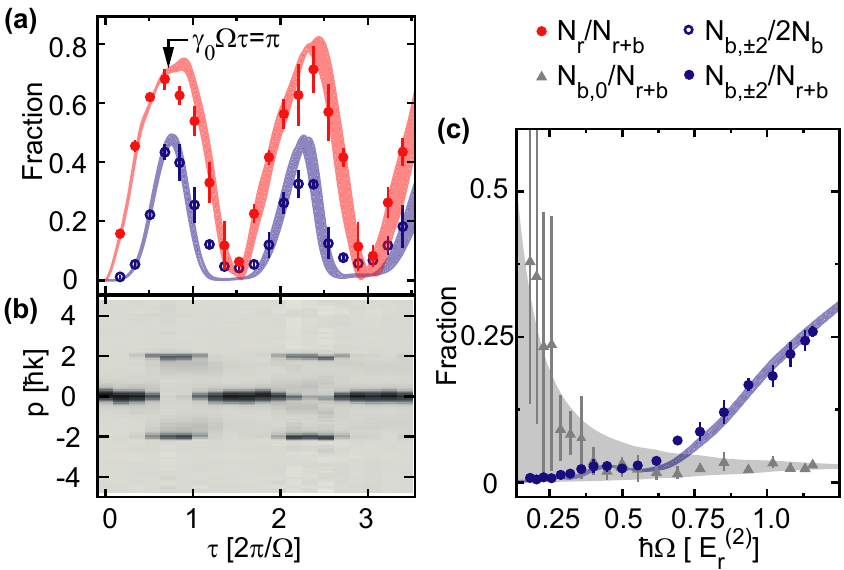}
\caption{(a) Resonant transfer to $\ket{0}$ and diffraction for $\Omega/2\pi=16.9$kHz. Shown are the transferred fraction (red filled points), and the diffracted fraction (blue open circles) in the momentum state $\ket{\pm2 \hbar k}$. The red and blue shaded regions are the results of a simulation including experimental uncertainties in $\delta$ and $\Omega$. The population in higher orders (not shown) is less than $2\%$ of the total number. (b) Observed diffraction patterns for $\ket{b}$ atoms.  Each line is a single TOF image summed over the direction perpendicular to the lattice. (c) Total diffraction signal as a function of $\Omega$ for $\gamma_0\Omega\tau = \pi$. The shaded blue (gray) area shows the band-structure prediction for atoms in the first diffracted order $N_{b,\pm2}/N_{r+b}$ (undiffracted fraction $N_{b,0}/N_{r+b}$) allowing for effective detunings up to $1.1$ kHz (see text).} \label{fig_2}
\end{figure}

Our experimental procedure, cf. Fig.~\ref{fig_1}a, uses microwave radiation to coherently couple an optically trapped $^{87}$Rb condensate (chemical potential $\mu=h\times1.0(1)$~kHz in a 60Hz trap at 1064nm \cite{Pertot2009}) in state $\ket{b}=\ket{F=1,m_F=-1}$  to the state $\ket{r}=\ket{2,-2}$ (separated from $\ket{b}$ by 6.8 GHz), which is exposed to a deep ($30 E_r$)  state-selective 1D optical lattice (wavelength $\lambda=2\pi/k=792$ nm, $\sigma^+$ polarization) not seen by the condensate \cite{Deutsch1998,Pertot2010}. We study population transfer into the lattice upon applying a rectangular microwave pulse, as a function of the applied Rabi frequency $\Omega\leq h \times 17$~kHz, the pulse duration $\tau$, and the detuning $\delta$ from the bare (i.e. no lattice) resonance between $\ket{b}$ and $\ket{r}$. In our experiments, residual magnetic-field fluctuations limit the shot-to-shot stability of the resonance to $\delta_Z/2\pi\sim0.6$~kHz; the typical uncertainty in determining $\Omega$ is $\sim 0.2$ kHz.

The lattice potential $V_r(z)=V_0 \sin^2{(k z)}$ modifies the microwave resonance condition, as $\ket{r}$ atoms are confined to lattice orbitals $\ket{n}$ (band index $n=0,1,2,...$). In our effectively blue-detuned lattice, the resonances are shifted upward by the zero-quasimomentum ($q$) band energies $\hbar\omega_{n}$. We characterize the lattice depth $V_0 = s E_r$ with $s=30(2)$ in terms of $E_r=\hbar\omega_r=(\hbar k)^2/2m = h\times 3.7~$kHz, where $m$ is the atomic mass (we also define $E_r^{(n)} = \hbar\omega_r^{(n)} = (n\hbar k)^2/2m$ used below). In the harmonic limit, $\hbar\omega_n\approx(n+\frac{1}{2})\hbar\omega_{ho}$, with $\hbar\omega_{ho}=2 \sqrt{s}E_r = h \times 40(1)$ kHz.

To study the orbital transfer, we first apply pulses of fixed $\tau = 60\mu$s, $\Omega/2\pi=8.1$kHz (corresponding to a $\pi$-pulse in the bare case) but variable detuning $\delta$. We then determine the orbital populations $P_n$ using a band-map sequence, consisting of a fast (1 ms) ramp-down of the lattice depth, subsequent free expansion (18 ms), Stern-Gerlach separation of $\ket{r}$ and $\ket{b}$, and absorption imaging. 
The observed transitions, cf. Fig.~\ref{fig_1}b, are in excellent agreement with the expectation from the lattice calibration, with line shapes that follow the theory for a Rabi pulse with a spectrally limited FWHM of 14 kHz. The free-space Rabi frequency ${\Omega=\bra{r} \boldsymbol{\mu} \cdot \textbf{B} \ket{b}/\hbar}$, is now modified by the orbital Franck-Condon overlap $\gamma_n=\langle n|\psi_0\rangle$ with the locally flat, single-particle condensate wavefunction $\ket{\psi_0}$; furthermore, parity conservation disallows transfer to the odd-$n$ orbitals.

We next study the coherent dynamics on resonance with the transition to the lowest orbital, $\ket{0}$ (where $\gamma_0=0.72$). For parameters similar to those above, Rabi-type oscillations can be observed for times up to $(\delta_Z)^{-1}\sim0.3$ ms. The first cycle for resonant coupling is shown in  Fig.~\ref{fig_1}c. Comparing the fitted frequency to $\Omega$ yields a reduction by $0.70(1)$, consistent with the calculated $\gamma_0$. We note that the oscillation amplitude is slighty reduced below unity, corresponding to a finite detuning of 1.5(1)~kHz in the Rabi model, which is within the systematic lattice calibration error (and comparable to differential mean-field shifts from the density compression in the lattice \cite{Pedri2001}).

However, the simple Rabi picture breaks down for stronger couplings. We find that, while the measured oscillation frequency increases with $\Omega$, the amplitude of the orbital population oscillation is strongly suppressed (by up to 30\%), cf. Fig.~\ref{fig_2}a.   At the same time, the  population in $\ket{b}$ undergoes high-contrast, pendell\"osung-type (albeit not number-conserving) oscillations between the initial momentum state $\ket{p=0\hbar k}$ and $\ket{p=\pm 2 \hbar k}$, cf. Fig.~\ref{fig_2}b, in phase with the population oscillation. As discussed below, the coupled internal-external dynamics is characteristic for the breakdown of the Born-Oppenheimer approximation.  The fraction of atoms in $\ket{0\hbar k}$ and in $\ket{\pm2\hbar k}$, for an effective $\pi$ pulse with $\gamma_0\Omega\tau=\pi$, are plotted in Fig.~\ref{fig_2}c. As a function of $\Omega$, the diffracted fraction increases rapidly once the coupling strength becomes comparable to $E_r^{(2)}$, while the population in $\ket{0 \hbar k}$ gets suppressed nearly completely, cf. Fig.~\ref{fig_2}c. We note that for small couplings $\Omega\lesssim5$kHz, a significant fraction of $\ket{b}$ atoms remain in $\ket{0\hbar k}$. However, this latter behavior is explained by the Rabi model if one allows for a small effective detuning of 1.1~kHz $\sim \delta_Z$.

To understand the appearance of additional momentum orders, one might naively describe the effect of extracting atoms from the condensate as that of an imaginary potential formed by the $q=0$ Bloch wave  (i.e. the array of ``absorbing'' Wannier functions). However, as in the case of a real periodic potential, the observed complete depletion of the initial state $\ket{0\hbar k}$ in this case would entail significantly populating at least one higher order $\ket{\pm2 j \hbar k}$ $(j=2,3,...)$, while the observed pendell\"osung dynamics involving $\ket{0,\pm2\hbar k}$ exists only in the perturbative limit of a weak optical potential \cite{Gadway2009}. 

\begin{figure}[t]
\centering
\includegraphics[width=\columnwidth]{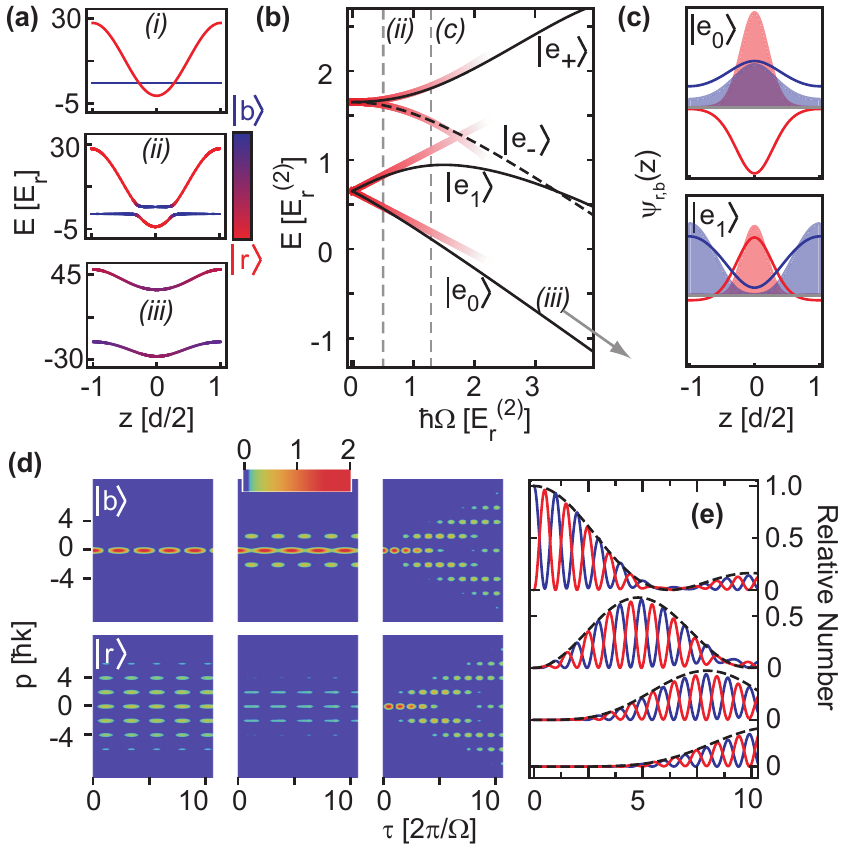}
\caption{Calculated dynamics from the nonadiabatic to the adiabatic regime. (a) (i-iii) Adiabatic potentials $\tilde{V}_i$ for coupling strengths $\hbar \Omega=\{0,1,15\} E_r^{(2)}$ and s=30. The position dependent internal state composition is encoded in color. (i) shows the bare potentials, while (iii) shows potentials in the adiabatic limit characterized by a near-constant dressed-state mixing angle. For weak coupling (ii), gaps are small and the mixing angle varies rapidly. (b) Nonadiabatic $q=0$ band energies for the four lowest states as a function of $\Omega$ for resonance with $\ket{n=0}$. The dashed line is a state of opposite parity that is decoupled from the other three states. (c) Spatial structure of the ground and first excited states in the bare-state basis, for $\hbar\Omega=1.25 E_r^{(2)}$. The shaded red and blue areas denote the square of the wave function, and the solid lines the wave function itself. (d) Diffractive dynamics in three regimes characterized by $\Omega/V_0 = 0.01,0.05,15$ with $V_0=100 E_r$, and (e) internal state dynamics in the case $\Omega/V_0 = 15$. The red and blue lines trace the total population in $\ket{r}$ and $\ket{b}$ while the dashed lines represent the expected envelopes for the $m=0,1,2,3$ (from top to bottom) diffraction orders for standard Kapitza-Dirac diffraction.}
\label{fig_3}
\end{figure}

A formal description of the system dynamics starts with the full rotating-wave Hamiltonian in the internal state basis $\{\ket{r},\ket{b} \}$ 
\begin{equation}
\mathcal{H}=\frac{\hat{p}^2}{2m} \textbf{I} +\frac{\hbar}{2} \begin{pmatrix} 2V_r(z)/\hbar-\delta & \Omega \\ \Omega & \delta \end{pmatrix} \equiv \mathcal{H}_{T}+ \mathcal{H}_V,
\label{hamiltonian}
\end{equation}
where $\hat{p}=-i\hbar \partial_z$ is the canonical momentum operator. In the usual approach, one then diagonalizes $\mathcal{H}_V$ in order to obtain the dressed states $\ket{\chi_i} = \alpha_i \ket{b} + \beta_i\ket{r}$ ($i=1,2$) and corresponding  optical potentials $\tilde{V}_i$, cf. Fig.~\ref{fig_3}a; the state of the system evolving under $\mathcal{H}$ is then expressed as $\Psi=\Sigma_j c_j\ket{\chi_j}$. However, this only works under the condition $\partial_z \ket{\chi_j} \ll \partial_z c_j$ (Born-Oppenheimer approximation)  \cite{Dalibard2011}. In our case (cf. Fig.~\ref{fig_3}a), where the bare-state energies cross at two positions within each lattice site and for weak coupling, (in our experiment, $\hbar\Omega\leq0.17 V_0$) the dressed-state mixing angle experiences rapid spatial variations.
Hence, the adiabatic condition is violated, and off-diagonal terms of the form $A_{ij}\propto \bra{\chi_i} \hat{p} \ket{\chi_j}$ become significant, leading to momentum-dependent mixing of the (adiabatic) dressed states.
With inseparable kinetic and potential terms in $\mathcal{H}$, the notion of an optical potential causing diffraction breaks down.

The results of a full diagonalization of the Hamiltonian $\mathcal{H}$ are shown in Fig.~\ref{fig_3}b. For the case of very weak microwave couplings $\hbar\Omega/E_r^{(2)}\ll1$ (where $ E_r^{(2)}/V_0<1)$, the energy levels are split by the two-level ac-Stark shift corresponding to the leading-order matrix element in the coupling.  For the energetically lowest states $\ket{e_0}$ and $\ket{e_1}$, the shifts are given by $\pm 0.72\hbar\Omega/2$, consistent with the Franck-Condon overlap $\gamma_0$ of the states $\ket{b,0 \hbar k}$ and $\ket{r,n=0}$. Switching on the microwave projects the wavefunction into a superposition between $\ket{e_0}$ and $\ket{e_1}$, whose beat note gives rise to the observed Rabi oscillations between $\ket{b, 0\hbar k}$ and $\ket{r,n=0}$.

The next-higher $\ket{e_{\pm}}$ states experience a shift due to off-resonant coupling of the symmetric (antisymmetric) momentum-state combinations $(\ket{b,2\hbar k}\pm\ket{b,-2\hbar k})/\sqrt{2}$ to $\ket{r,n=0}$ ($\ket{r,n=1}$). For intermediate couplings $\hbar\Omega\sim E_r^{(2)}$, the admixture of $(\ket{b,2\hbar k}+\ket{b,-2\hbar k})/\sqrt{2}$ to $\ket{e_0}$ and $\ket{e_1}$ induces a downward curving of their energies and gives them the periodicity of the lattice, cf. Fig.~\ref{fig_3}c. In this regime, the beat between $\ket{e_0}$ and $\ket{e_1}$ leads to the dynamical formation of a periodic density modulation, in phase with the internal-state evolution. While at $t=0$ the sum of these states yields the unmodulated state $\ket{b,0 \hbar k}$,  at $\Omega t\approx \pi/(\gamma_0)$ they acquire a  $\pi$ phase shift and add to $(\ket{b,2\hbar k} + \ket{b,-2\hbar k})/\sqrt{2}$ , which produces the observed diffraction pattern (with a small contamination by $\ket{b,0 \hbar k}$ due to the admixture of $\ket{e_{+}}$).

Strong couplings $\hbar\Omega\gg E_r^{(2)}$ induce the admixture of higher-order momentum-state combinations, and since $E_r^{(2)}\sim\hbar\omega_{ho}$ (for $s\sim10$) the dynamics also involves higher even$-n$ orbitals. Taken together, the internal and external dynamics eventually decouple: the former proceeds as a fast oscillation between $\ket{b}$ and $\ket{r}$ (with Rabi frequency $\Omega$), while in the rotating frame the adiabatic dressed states $\ket{\chi_{1,2}} = (\ket{b}\pm\ket{r})/\sqrt{2}$ independently undergo diffraction in the half-depth optical potentials $V_{1,2} = V(z)/2\pm\hbar\Omega/2$, with populations $P_n(\tau) = J_n^2[V_0\tau/4\hbar]$ spread over many orders $n$ as expected for standard Kapitza-Dirac diffraction, cf. Fig.~\ref{fig_3}d and \ref{fig_3}e. Indeed, for a $\pi$-pulse with $\tau\Omega\sim\pi$, the condition $\Omega\gg\omega_{ho}$ reproduces the Raman-Nath criterion  $\tau\omega_{ho}\ll1$  \cite{Ovchinnikov1999} (we caution that reaching this adiabatic limit would require increasing the microwave power by three orders of magnitude, which is outside our technical capabilities).

Our model accurately reproduces all the observed dynamics in Fig.~\ref{fig_2}. The narrowing of the diffraction signal in Fig.~\ref{fig_2}a (as compared to the orbital transfer) is easily explained by considering that the signal is  more pronounced with the depletion of the background in $\ket{b, 0\hbar k}$.

\begin{figure}[t]
\centering
\includegraphics[width=\columnwidth]{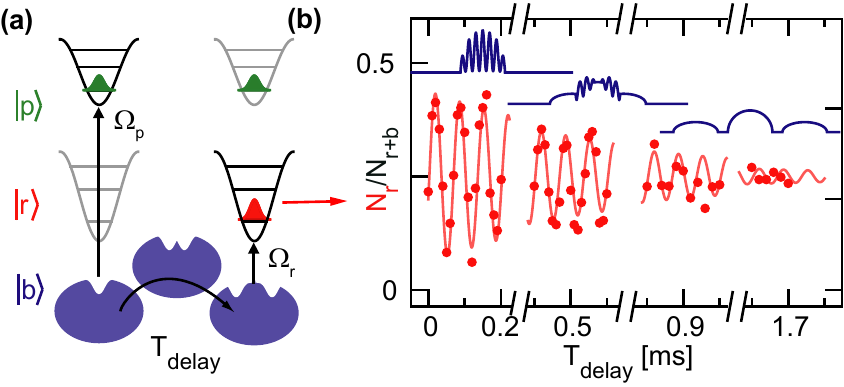}
\caption{In-situ probing of matter-wave interference. (a) A pump pulse ($\gamma_0\Omega\tau=2\pi/3$, $\tau=40\mu$s) creates excitations in the condensate by removing atoms into a second trapped state $\ket{p}\equiv\ket{2,-1}$, and is followed by free evolution of the condensate density profile for time $T_\mathrm{delay}$ before a probe pulse ($\gamma_0\Omega\tau=2\pi/5$, $\tau=14\mu$s) transfers atoms into the trapped state $\ket{r}$. (b) Probe pulse yield as a function of delay time, displaying oscillations at the recoil frequency. The solid line is a sinusoidal fit with frequency $\omega/2\pi = 15.0~$kHz, and exponential damping time $\tilde{\tau} = 0.66$ms, corresponding to the beat note and characteristic time for the  wavepackets separating after the probe pulse. The insets illustrate the evolving wave-packet interference after application of the pump pulse.} \label{fig_4}
\end{figure}

As a simple application of our orbital transfer scheme going beyond mere diffraction, we implement an in-situ detector for matter-wave interferometry, in which the initially unpopulated lattice sites act as an array of probes for the condensate, cf. Fig.~\ref{fig_4}a. The orbital transfer can be viewed as (microwave-induced) local tunneling of atoms into individual lattice sites. As the transferred atoms are detected, the operation is thus reminiscent of detecting electrons in a tunneling microscope.
We first imprint a modulation into the condensate (``pump"), again using orbital transfer. To keep this step separate from the subsequent detection (``probe"), a third internal state $\ket{p}\equiv\ket{2,-1}$ with a lattice potential $V_p = (2/3) V_r$ is used, precluding any unwanted coherence effects in detection. Figure \ref{fig_4}b shows the resulting probe signal $N_r$ as a function of the time delay $T_\textrm{delay}$ between the two pulses. We observe an oscillation in $N_r$, consistent with an on-site beat note at $\omega_r^{(2)}$, generated by matter wave interference between recoiling $\ket{b,\pm2\hbar k}$ and stationary $\ket{b,0\hbar k}$ packets. The decaying envelope reflects a gradual loss of spatial overlap between the wavepackets. Compared to Kapitza-Dirac-based matter-wave interferometry schemes \cite{Gupta2002,Campbell2005}, our 2-pulse sequence allows for a direct, in-situ detection of the matter wave interference contrast, without the need for an additional probe laser.

In the present work, we have studied the dynamics of a condensate coupled to the orbitals of an array of lattice sites.  We point out that we do not observe effects of the collisional interaction $\mu$ (such as a modification of the ground-state wavefunctions \cite{Abad2013}), as $\mu$ is on the order of fluctuations of the orbital-resonance frequency. In the interacting regime $\hbar\Omega<\mu$, the healing length $\xi=\hbar/(2m\mu)^{-1/2}$ is compatible with phonon excitation, at a wavelength $\lambdabar_s = (\mu/m)^{1/2}/\Omega = (\mu/\hbar\Omega) \xi > \xi$. The coupling of the single-site dynamics to phonons should allow for the study of dissipative phenomena \cite{Recati2005}, including dissipative quantum phase transitions driven by phonon-mediated coupling between sites \cite{Orth2008}. Entering this regime in our setup requires a reduction of the ambient-magnetic field fluctuations below $100\mu$G, which is reachable with present-day technology.

In summary, we have demonstrated a novel regime for the diffraction of weakly-dressed matter-waves from a standing wave of light. In this regime, nonadiabatic transitions induce a strong coupling between the internal and external degrees of freedom, such that a description of the standing wave through an optical potential is not possible. In the limit of strong microwave coupling, the decoupling of the internal and external dynamics results in conventional Kapitza-Dirac diffraction from an optical potential. While the coherent internal dynamics give rise to particle-like excitations in the regime explored in this work, an extension into the regime of even weaker microwave coupling, in which the superfluid excitations become sound-like, should prove useful in the study of dissipative phenomena.

\begin{acknowledgements}
We acknowledge funding from NSF (PHY-1205894). JR and MS gratefully acknowledge a GAANN fellowship from the DoEd. AP acknowledges support by a grant from ESPOL-SENESCYT.  We thank K. Le Hur and P. Orth for stimulating discussions, and T. Bergeman and M. G. Cohen for a careful reading of the manuscript.
\end{acknowledgements}

\bibliography{diffraction.bib}

\begin{thebibliography}{33}
\expandafter\ifx\csname natexlab\endcsname\relax\def\natexlab#1{#1}\fi
\expandafter\ifx\csname bibnamefont\endcsname\relax
  \def\bibnamefont#1{#1}\fi
\expandafter\ifx\csname bibfnamefont\endcsname\relax
  \def\bibfnamefont#1{#1}\fi
\expandafter\ifx\csname citenamefont\endcsname\relax
  \def\citenamefont#1{#1}\fi
\expandafter\ifx\csname url\endcsname\relax
  \def\url#1{\texttt{#1}}\fi
\expandafter\ifx\csname urlprefix\endcsname\relax\def\urlprefix{URL }\fi
\providecommand{\bibinfo}[2]{#2}
\providecommand{\eprint}[2][]{\url{#2}}

\bibitem[{\citenamefont{Bragg and Bragg}(1913)}]{Bragg1913}
\bibinfo{author}{\bibfnamefont{W.~H.} \bibnamefont{Bragg}} \bibnamefont{and}
  \bibinfo{author}{\bibfnamefont{W.~L.} \bibnamefont{Bragg}},
  \bibinfo{journal}{Proc. R. Soc. Lond. A} \textbf{\bibinfo{volume}{88}},
  \bibinfo{pages}{428} (\bibinfo{year}{1913}).

\bibitem[{\citenamefont{Laue}(1913)}]{Laue1913}
\bibinfo{author}{\bibfnamefont{M.}~\bibnamefont{Laue}},
  \bibinfo{journal}{Physik. Z.} \textbf{\bibinfo{volume}{14}},
  \bibinfo{pages}{421} (\bibinfo{year}{1913}).

\bibitem[{\citenamefont{Davisson and Germer}(1927)}]{Davisson1927}
\bibinfo{author}{\bibfnamefont{C.}~\bibnamefont{Davisson}} \bibnamefont{and}
  \bibinfo{author}{\bibfnamefont{L.~H.} \bibnamefont{Germer}},
  \bibinfo{journal}{Nature} \textbf{\bibinfo{volume}{119}},
  \bibinfo{pages}{558} (\bibinfo{year}{1927}).

\bibitem[{\citenamefont{Meystre}(2001)}]{Meystre2001}
\bibinfo{author}{\bibfnamefont{P.}~\bibnamefont{Meystre}},
  \emph{\bibinfo{title}{Atom Optics}} (\bibinfo{publisher}{Springer},
  \bibinfo{year}{2001}).

\bibitem[{\citenamefont{Cronin et~al.}(2009)\citenamefont{Cronin, Schmiedmayer,
  and Pritchard}}]{Cronin2009}
\bibinfo{author}{\bibfnamefont{A.~D.} \bibnamefont{Cronin}},
  \bibinfo{author}{\bibfnamefont{J.}~\bibnamefont{Schmiedmayer}},
  \bibnamefont{and} \bibinfo{author}{\bibfnamefont{D.~E.}
  \bibnamefont{Pritchard}}, \bibinfo{journal}{Rev. Mod. Phys.}
  \textbf{\bibinfo{volume}{81}}, \bibinfo{pages}{1051} (\bibinfo{year}{2009}).

\bibitem[{\citenamefont{Kapitza and Dirac}(1933)}]{Kapitza1933}
\bibinfo{author}{\bibfnamefont{P.~L.} \bibnamefont{Kapitza}} \bibnamefont{and}
  \bibinfo{author}{\bibfnamefont{P.~A.~M.} \bibnamefont{Dirac}},
  \bibinfo{journal}{Math. Proc. Cambridge} \textbf{\bibinfo{volume}{29}},
  \bibinfo{pages}{297} (\bibinfo{year}{1933}).

\bibitem[{\citenamefont{Freimund et~al.}(2001)\citenamefont{Freimund,
  Aflatooni, and Batelaan}}]{Freimund2001}
\bibinfo{author}{\bibfnamefont{D.~L.} \bibnamefont{Freimund}},
  \bibinfo{author}{\bibfnamefont{K.}~\bibnamefont{Aflatooni}},
  \bibnamefont{and} \bibinfo{author}{\bibfnamefont{H.}~\bibnamefont{Batelaan}},
  \bibinfo{journal}{Nature} \textbf{\bibinfo{volume}{413}},
  \bibinfo{pages}{142} (\bibinfo{year}{2001}).

\bibitem[{\citenamefont{Gould et~al.}(1986)\citenamefont{Gould, Ruff, and
  Pritchard}}]{Gould1986}
\bibinfo{author}{\bibfnamefont{P.~L.} \bibnamefont{Gould}},
  \bibinfo{author}{\bibfnamefont{G.~A.} \bibnamefont{Ruff}}, \bibnamefont{and}
  \bibinfo{author}{\bibfnamefont{D.~E.} \bibnamefont{Pritchard}},
  \bibinfo{journal}{Phys. Rev. Lett.} \textbf{\bibinfo{volume}{56}},
  \bibinfo{pages}{827} (\bibinfo{year}{1986}).

\bibitem[{\citenamefont{Ovchinnikov et~al.}(1999)\citenamefont{Ovchinnikov,
  M\"{u}ller, Doery, Vredenbregt, Helmerson, Rolston, and
  Phillips}}]{Ovchinnikov1999}
\bibinfo{author}{\bibfnamefont{Y.~B.} \bibnamefont{Ovchinnikov}},
  \bibinfo{author}{\bibfnamefont{J.~H.} \bibnamefont{M\"{u}ller}},
  \bibinfo{author}{\bibfnamefont{M.~R.} \bibnamefont{Doery}},
  \bibinfo{author}{\bibfnamefont{E.~J.~D.} \bibnamefont{Vredenbregt}},
  \bibinfo{author}{\bibfnamefont{K.}~\bibnamefont{Helmerson}},
  \bibinfo{author}{\bibfnamefont{S.~L.} \bibnamefont{Rolston}},
  \bibnamefont{and} \bibinfo{author}{\bibfnamefont{W.~D.}
  \bibnamefont{Phillips}}, \bibinfo{journal}{Phys. Rev. Lett.}
  \textbf{\bibinfo{volume}{83}}, \bibinfo{pages}{284} (\bibinfo{year}{1999}).

\bibitem[{\citenamefont{Stenger et~al.}(1999)\citenamefont{Stenger, Inouye,
  Chikkatur, Stamper-Kurn, Pritchard, and Ketterle}}]{Stenger1999}
\bibinfo{author}{\bibfnamefont{J.}~\bibnamefont{Stenger}},
  \bibinfo{author}{\bibfnamefont{S.}~\bibnamefont{Inouye}},
  \bibinfo{author}{\bibfnamefont{A.~P.} \bibnamefont{Chikkatur}},
  \bibinfo{author}{\bibfnamefont{D.~M.} \bibnamefont{Stamper-Kurn}},
  \bibinfo{author}{\bibfnamefont{D.~E.} \bibnamefont{Pritchard}},
  \bibnamefont{and} \bibinfo{author}{\bibfnamefont{W.}~\bibnamefont{Ketterle}},
  \bibinfo{journal}{Phys. Rev. Lett.} \textbf{\bibinfo{volume}{82}},
  \bibinfo{pages}{4569} (\bibinfo{year}{1999}).

\bibitem[{\citenamefont{Ketterle and Druten}(1996)}]{Ketterle1996}
\bibinfo{author}{\bibfnamefont{W.}~\bibnamefont{Ketterle}} \bibnamefont{and}
  \bibinfo{author}{\bibfnamefont{N.~J.~V.} \bibnamefont{Druten}},
  \bibinfo{journal}{Adv. At., Mol., Opt. Phys.} \textbf{\bibinfo{volume}{37}},
  \bibinfo{pages}{181} (\bibinfo{year}{1996}).

\bibitem[{\citenamefont{Zobay and Garraway}(2001)}]{Zobay2001}
\bibinfo{author}{\bibfnamefont{O.}~\bibnamefont{Zobay}} \bibnamefont{and}
  \bibinfo{author}{\bibfnamefont{B.~M.} \bibnamefont{Garraway}},
  \bibinfo{journal}{Phys. Rev. Lett.} \textbf{\bibinfo{volume}{86}},
  \bibinfo{pages}{1195} (\bibinfo{year}{2001}).

\bibitem[{\citenamefont{Colombe et~al.}(2004)\citenamefont{Colombe, Knyazchyan,
  Morizot, Mercier, Lorent, and Perrin}}]{Colombe2004}
\bibinfo{author}{\bibfnamefont{Y.}~\bibnamefont{Colombe}},
  \bibinfo{author}{\bibfnamefont{E.}~\bibnamefont{Knyazchyan}},
  \bibinfo{author}{\bibfnamefont{O.}~\bibnamefont{Morizot}},
  \bibinfo{author}{\bibfnamefont{B.}~\bibnamefont{Mercier}},
  \bibinfo{author}{\bibfnamefont{V.}~\bibnamefont{Lorent}}, \bibnamefont{and}
  \bibinfo{author}{\bibfnamefont{H.}~\bibnamefont{Perrin}},
  \bibinfo{journal}{Europhys. Lett.} \textbf{\bibinfo{volume}{67}},
  \bibinfo{pages}{593} (\bibinfo{year}{2004}).

\bibitem[{\citenamefont{Cohen-Tannoudji
  et~al.}(2004)\citenamefont{Cohen-Tannoudji, Dupont-Roc, and
  Grynberg}}]{Cohen-Tannoudji2004}
\bibinfo{author}{\bibfnamefont{C.}~\bibnamefont{Cohen-Tannoudji}},
  \bibinfo{author}{\bibfnamefont{J.}~\bibnamefont{Dupont-Roc}},
  \bibnamefont{and} \bibinfo{author}{\bibfnamefont{G.}~\bibnamefont{Grynberg}},
  \emph{\bibinfo{title}{Atom Photon Interactions}}
  (\bibinfo{publisher}{WILEY-VHC}, \bibinfo{year}{2004}).

\bibitem[{\citenamefont{Lesanovsky et~al.}(2006)\citenamefont{Lesanovsky,
  Schumm, Hofferberth, Andersson, Kr\"{u}ger, and
  Schmiedmayer}}]{Lesanovsky2006}
\bibinfo{author}{\bibfnamefont{I.}~\bibnamefont{Lesanovsky}},
  \bibinfo{author}{\bibfnamefont{T.}~\bibnamefont{Schumm}},
  \bibinfo{author}{\bibfnamefont{S.}~\bibnamefont{Hofferberth}},
  \bibinfo{author}{\bibfnamefont{L.~M.} \bibnamefont{Andersson}},
  \bibinfo{author}{\bibfnamefont{P.}~\bibnamefont{Kr\"{u}ger}},
  \bibnamefont{and}
  \bibinfo{author}{\bibfnamefont{J.}~\bibnamefont{Schmiedmayer}},
  \bibinfo{journal}{Phys. Rev. A} \textbf{\bibinfo{volume}{73}},
  \bibinfo{pages}{033619} (\bibinfo{year}{2006}).

\bibitem[{\citenamefont{White et~al.}(2006)\citenamefont{White, Gao, Pasienski,
  and DeMarco}}]{White2006}
\bibinfo{author}{\bibfnamefont{M.}~\bibnamefont{White}},
  \bibinfo{author}{\bibfnamefont{H.}~\bibnamefont{Gao}},
  \bibinfo{author}{\bibfnamefont{M.}~\bibnamefont{Pasienski}},
  \bibnamefont{and} \bibinfo{author}{\bibfnamefont{B.}~\bibnamefont{DeMarco}},
  \bibinfo{journal}{Phys. Rev. A} \textbf{\bibinfo{volume}{74}},
  \bibinfo{pages}{023616} (\bibinfo{year}{2006}).

\bibitem[{\citenamefont{Hofferberth et~al.}(2006)\citenamefont{Hofferberth,
  Lesanovsky, Fischer, Verdu, and Schmiedmayer}}]{Hofferberth2006}
\bibinfo{author}{\bibfnamefont{S.}~\bibnamefont{Hofferberth}},
  \bibinfo{author}{\bibfnamefont{I.}~\bibnamefont{Lesanovsky}},
  \bibinfo{author}{\bibfnamefont{B.}~\bibnamefont{Fischer}},
  \bibinfo{author}{\bibfnamefont{J.}~\bibnamefont{Verdu}}, \bibnamefont{and}
  \bibinfo{author}{\bibfnamefont{J.}~\bibnamefont{Schmiedmayer}},
  \bibinfo{journal}{Nature Phys.} \textbf{\bibinfo{volume}{2}},
  \bibinfo{pages}{710} (\bibinfo{year}{2006}).

\bibitem[{\citenamefont{Lundblad et~al.}(2008)\citenamefont{Lundblad, Lee,
  Spielman, Brown, Phillips, and Porto}}]{Lundblad2008}
\bibinfo{author}{\bibfnamefont{N.}~\bibnamefont{Lundblad}},
  \bibinfo{author}{\bibfnamefont{P.~J.} \bibnamefont{Lee}},
  \bibinfo{author}{\bibfnamefont{I.~B.} \bibnamefont{Spielman}},
  \bibinfo{author}{\bibfnamefont{B.~L.} \bibnamefont{Brown}},
  \bibinfo{author}{\bibfnamefont{W.~D.} \bibnamefont{Phillips}},
  \bibnamefont{and} \bibinfo{author}{\bibfnamefont{J.~V.} \bibnamefont{Porto}},
  \bibinfo{journal}{Phys. Rev. Lett.} \textbf{\bibinfo{volume}{100}},
  \bibinfo{pages}{150401} (\bibinfo{year}{2008}).

\bibitem[{\citenamefont{Lin et~al.}(2009)\citenamefont{Lin, Compton,
  Jimenez-Garcia, Porto, and Spielman}}]{Lin2009}
\bibinfo{author}{\bibfnamefont{Y.~J.} \bibnamefont{Lin}},
  \bibinfo{author}{\bibfnamefont{R.~L.} \bibnamefont{Compton}},
  \bibinfo{author}{\bibfnamefont{K.}~\bibnamefont{Jimenez-Garcia}},
  \bibinfo{author}{\bibfnamefont{J.~V.} \bibnamefont{Porto}}, \bibnamefont{and}
  \bibinfo{author}{\bibfnamefont{I.~B.} \bibnamefont{Spielman}},
  \bibinfo{journal}{Nature} \textbf{\bibinfo{volume}{462}},
  \bibinfo{pages}{628} (\bibinfo{year}{2009}).

\bibitem[{\citenamefont{Yi et~al.}(2008)\citenamefont{Yi, Daley, Pupillo, and
  Zoller}}]{Yi2008}
\bibinfo{author}{\bibfnamefont{W.}~\bibnamefont{Yi}},
  \bibinfo{author}{\bibfnamefont{A.~J.} \bibnamefont{Daley}},
  \bibinfo{author}{\bibfnamefont{G.}~\bibnamefont{Pupillo}}, \bibnamefont{and}
  \bibinfo{author}{\bibfnamefont{P.}~\bibnamefont{Zoller}},
  \bibinfo{journal}{New J. Phys.} \textbf{\bibinfo{volume}{10}},
  \bibinfo{pages}{073015} (\bibinfo{year}{2008}).

\bibitem[{\citenamefont{Lundblad et~al.}(2014)\citenamefont{Lundblad, Ansari,
  Guo, and Moan}}]{Lundblad2014}
\bibinfo{author}{\bibfnamefont{N.}~\bibnamefont{Lundblad}},
  \bibinfo{author}{\bibfnamefont{S.}~\bibnamefont{Ansari}},
  \bibinfo{author}{\bibfnamefont{Y.}~\bibnamefont{Guo}}, \bibnamefont{and}
  \bibinfo{author}{\bibfnamefont{E.}~\bibnamefont{Moan}},
  \bibinfo{journal}{Phys. Rev. A} \textbf{\bibinfo{volume}{90}},
  \bibinfo{pages}{053612} (\bibinfo{year}{2014}).

\bibitem[{\citenamefont{Dalibard et~al.}(2011)\citenamefont{Dalibard, Gerbier,
  Juzeli\={u}nas, and \"{O}hberg}}]{Dalibard2011}
\bibinfo{author}{\bibfnamefont{J.}~\bibnamefont{Dalibard}},
  \bibinfo{author}{\bibfnamefont{F.}~\bibnamefont{Gerbier}},
  \bibinfo{author}{\bibfnamefont{G.}~\bibnamefont{Juzeli\={u}nas}},
  \bibnamefont{and}
  \bibinfo{author}{\bibfnamefont{P.}~\bibnamefont{\"{O}hberg}},
  \bibinfo{journal}{Rev. Mod. Phys.} \textbf{\bibinfo{volume}{83}},
  \bibinfo{pages}{1523} (\bibinfo{year}{2011}).

\bibitem[{\citenamefont{Pertot et~al.}(2009)\citenamefont{Pertot, Greif,
  Albert, Gadway, and Schneble}}]{Pertot2009}
\bibinfo{author}{\bibfnamefont{D.}~\bibnamefont{Pertot}},
  \bibinfo{author}{\bibfnamefont{D.}~\bibnamefont{Greif}},
  \bibinfo{author}{\bibfnamefont{S.}~\bibnamefont{Albert}},
  \bibinfo{author}{\bibfnamefont{B.}~\bibnamefont{Gadway}}, \bibnamefont{and}
  \bibinfo{author}{\bibfnamefont{D.}~\bibnamefont{Schneble}},
  \bibinfo{journal}{J. Phys. B: At., Mol. Opt. Phys.}
  \textbf{\bibinfo{volume}{42}}, \bibinfo{pages}{215305}
  (\bibinfo{year}{2009}).

\bibitem[{\citenamefont{Deutsch and Jessen}(1998)}]{Deutsch1998}
\bibinfo{author}{\bibfnamefont{I.~H.} \bibnamefont{Deutsch}} \bibnamefont{and}
  \bibinfo{author}{\bibfnamefont{P.~S.} \bibnamefont{Jessen}},
  \bibinfo{journal}{Phys. Rev. A} \textbf{\bibinfo{volume}{57}},
  \bibinfo{pages}{1972} (\bibinfo{year}{1998}).

\bibitem[{\citenamefont{Pertot et~al.}(2010)\citenamefont{Pertot, Gadway, and
  Schneble}}]{Pertot2010}
\bibinfo{author}{\bibfnamefont{D.}~\bibnamefont{Pertot}},
  \bibinfo{author}{\bibfnamefont{B.}~\bibnamefont{Gadway}}, \bibnamefont{and}
  \bibinfo{author}{\bibfnamefont{D.}~\bibnamefont{Schneble}},
  \bibinfo{journal}{Phys. Rev. Lett.} \textbf{\bibinfo{volume}{104}},
  \bibinfo{pages}{200402} (\bibinfo{year}{2010}).

\bibitem[{\citenamefont{Gadway et~al.}(2012)\citenamefont{Gadway, Pertot,
  Reeves, and Schneble}}]{Gadway2012}
\bibinfo{author}{\bibfnamefont{B.}~\bibnamefont{Gadway}},
  \bibinfo{author}{\bibfnamefont{D.}~\bibnamefont{Pertot}},
  \bibinfo{author}{\bibfnamefont{J.}~\bibnamefont{Reeves}}, \bibnamefont{and}
  \bibinfo{author}{\bibfnamefont{D.}~\bibnamefont{Schneble}},
  \bibinfo{journal}{Nature Phys.} \textbf{\bibinfo{volume}{8}},
  \bibinfo{pages}{544} (\bibinfo{year}{2012}).

\bibitem[{\citenamefont{Pedri et~al.}(2001)\citenamefont{Pedri, Pitaevskii,
  Stringari, Fort, Burger, Cataliotti, Maddaloni, Minardi, and
  Inguscio}}]{Pedri2001}
\bibinfo{author}{\bibfnamefont{P.}~\bibnamefont{Pedri}},
  \bibinfo{author}{\bibfnamefont{L.}~\bibnamefont{Pitaevskii}},
  \bibinfo{author}{\bibfnamefont{S.}~\bibnamefont{Stringari}},
  \bibinfo{author}{\bibfnamefont{C.}~\bibnamefont{Fort}},
  \bibinfo{author}{\bibfnamefont{S.}~\bibnamefont{Burger}},
  \bibinfo{author}{\bibfnamefont{F.~S.} \bibnamefont{Cataliotti}},
  \bibinfo{author}{\bibfnamefont{P.}~\bibnamefont{Maddaloni}},
  \bibinfo{author}{\bibfnamefont{F.}~\bibnamefont{Minardi}}, \bibnamefont{and}
  \bibinfo{author}{\bibfnamefont{M.}~\bibnamefont{Inguscio}},
  \bibinfo{journal}{Phys. Rev. Lett.} \textbf{\bibinfo{volume}{87}},
  \bibinfo{pages}{220401} (\bibinfo{year}{2001}).

\bibitem[{\citenamefont{Gadway et~al.}(2009)\citenamefont{Gadway, Pertot,
  Reimann, Cohen, and Schneble}}]{Gadway2009}
\bibinfo{author}{\bibfnamefont{B.}~\bibnamefont{Gadway}},
  \bibinfo{author}{\bibfnamefont{D.}~\bibnamefont{Pertot}},
  \bibinfo{author}{\bibfnamefont{R.}~\bibnamefont{Reimann}},
  \bibinfo{author}{\bibfnamefont{M.~G.} \bibnamefont{Cohen}}, \bibnamefont{and}
  \bibinfo{author}{\bibfnamefont{D.}~\bibnamefont{Schneble}},
  \bibinfo{journal}{Opt. Express} \textbf{\bibinfo{volume}{17}},
  \bibinfo{pages}{19173} (\bibinfo{year}{2009}).

\bibitem[{\citenamefont{Gupta et~al.}(2002)\citenamefont{Gupta, Dieckmann,
  Hadzibabic, and Pritchard}}]{Gupta2002}
\bibinfo{author}{\bibfnamefont{S.}~\bibnamefont{Gupta}},
  \bibinfo{author}{\bibfnamefont{K.}~\bibnamefont{Dieckmann}},
  \bibinfo{author}{\bibfnamefont{Z.}~\bibnamefont{Hadzibabic}},
  \bibnamefont{and} \bibinfo{author}{\bibfnamefont{D.~E.}
  \bibnamefont{Pritchard}}, \bibinfo{journal}{Phys. Rev. Lett.}
  \textbf{\bibinfo{volume}{89}}, \bibinfo{pages}{140401}
  (\bibinfo{year}{2002}).

\bibitem[{\citenamefont{Campbell et~al.}(2005)\citenamefont{Campbell,
  Leanhardt, Mun, Boyd, Streed, Ketterle, and Pritchard}}]{Campbell2005}
\bibinfo{author}{\bibfnamefont{G.~K.} \bibnamefont{Campbell}},
  \bibinfo{author}{\bibfnamefont{A.~E.} \bibnamefont{Leanhardt}},
  \bibinfo{author}{\bibfnamefont{J.}~\bibnamefont{Mun}},
  \bibinfo{author}{\bibfnamefont{M.}~\bibnamefont{Boyd}},
  \bibinfo{author}{\bibfnamefont{E.~W.} \bibnamefont{Streed}},
  \bibinfo{author}{\bibfnamefont{W.}~\bibnamefont{Ketterle}}, \bibnamefont{and}
  \bibinfo{author}{\bibfnamefont{D.~E.} \bibnamefont{Pritchard}},
  \bibinfo{journal}{Phys. Rev. Lett.} \textbf{\bibinfo{volume}{94}},
  \bibinfo{pages}{170403} (\bibinfo{year}{2005}).

\bibitem[{\citenamefont{Abad and Recati}(2013)}]{Abad2013}
\bibinfo{author}{\bibfnamefont{M.}~\bibnamefont{Abad}} \bibnamefont{and}
  \bibinfo{author}{\bibfnamefont{A.}~\bibnamefont{Recati}},
  \bibinfo{journal}{EPJ D} \textbf{\bibinfo{volume}{67}}, \bibinfo{pages}{1}
  (\bibinfo{year}{2013}).

\bibitem[{\citenamefont{Recati et~al.}(2005)\citenamefont{Recati, Fedichev,
  Zwerger, von Delft, and Zoller}}]{Recati2005}
\bibinfo{author}{\bibfnamefont{A.}~\bibnamefont{Recati}},
  \bibinfo{author}{\bibfnamefont{P.~O.} \bibnamefont{Fedichev}},
  \bibinfo{author}{\bibfnamefont{W.}~\bibnamefont{Zwerger}},
  \bibinfo{author}{\bibfnamefont{J.}~\bibnamefont{von Delft}},
  \bibnamefont{and} \bibinfo{author}{\bibfnamefont{P.}~\bibnamefont{Zoller}},
  \bibinfo{journal}{Phys. Rev. Lett.} \textbf{\bibinfo{volume}{94}},
  \bibinfo{pages}{040404} (\bibinfo{year}{2005}).

\bibitem[{\citenamefont{Orth et~al.}(2008)\citenamefont{Orth, Stanic, and
  Le~Hur}}]{Orth2008}
\bibinfo{author}{\bibfnamefont{P.~P.} \bibnamefont{Orth}},
  \bibinfo{author}{\bibfnamefont{I.}~\bibnamefont{Stanic}}, \bibnamefont{and}
  \bibinfo{author}{\bibfnamefont{K.}~\bibnamefont{Le~Hur}},
  \bibinfo{journal}{Phys. Rev. A} \textbf{\bibinfo{volume}{77}},
  \bibinfo{pages}{051601} (\bibinfo{year}{2008}).

\end{thebibliography}

\end{document}